\documentstyle[aaspp4,12pt]{article}
 \def\simlt{\lower.5ex\hbox{$\; \buildrel < \over \sim \;$}}
  \def\simgt{\lower.5ex\hbox{$\; \buildrel > \over \sim \;$}}
\begin{document}

\title{Gravitational Magnification of the Cosmic Microwave Background}

\author{R. Benton Metcalf and Joseph Silk}
\affil{Departments of Physics  and Astronomy, and Center for Particle
Astrophysics \\ University of California, Berkeley, California 84720}

\begin{abstract}
Some aspects of gravitational lensing by large scale structure (LSS) are investigated.  
We show that lensing causes the damping tail of the cosmic microwave background (CMB) 
power spectrum to fall less rapidly with decreasing angular scale than previously expected.  
This is due to a transfer of power from larger to smaller angular scales which produces a 
fractional change in power spectrum that increases rapidly beyond $\ell \sim 2000$.  We also 
find that lensing produces a nonzero mean magnification of structures on  surfaces of constant 
redshift if weighted by area on the sky.  This is a result of the fact that light-rays that are 
evenly distributed on the sky oversample overdense regions.  However this mean magnification 
has a negligible affect on the CMB power spectrum.  A new expression for the lensed power 
spectrum is derived and it is found that future precision observations the high-$\ell$ 
tail of the power spectrum will need to take into account lensing when determining cosmological 
parameters.
\end{abstract}

\section{Introduction}

Previous discussions of gravitational lensing by large-scale structure have concentrated on 
calculating the shear and convergence along unperturbed light-paths, i.e. what the geodesics 
would be were there no fluctuations (e.g. \cite{seljak}, and references cited therein).  Three basic 
methods have been adopted.  The first is by numerical simulation (e.g. \cite{fuk94}). 
This method often suffers from limited resolution and overly idealized cosmological models.  Another 
method has been to use a model where light 
travels freely in a constant background density between clumps of localized 
mass densities \cite{fuk94,bess94}.  This is not considered to be a realistic cosmological model,
because of the wide range of length scales on which galaxy clustering is 
observed.  What appears to be the  best method thus far is to take a smooth field of 
density fluctuations and calculate the shear and convergence along unperturbed 
light-paths.  This can be done with the use of optical scalars \cite{gunn67,blan91} or equivalently by 
using methods based on those of \cite{kais92}.

In particular, \cite{seljak} has applied the techniques of Kaiser \cite{kais92} to the 
lensing of the CMB.  He found that lensing results in  a relatively small smoothing of the 
CMB power spectrum which makes peaks and troughs somewhat less distinct.  This smoothing is 
due to fluctuations in the magnification of structures on the surface of last scattering.  
The average magnification was assumed to be zero, as it is to first order.  Seljak also found that 
evolving the deflecting density fluctuations by linear or nonlinear theory makes little difference 
in the results for $\ell < 1000$.

We show here that deviations of the light-paths from their form in an 
unperturbed universe result not only in fluctuations in the magnification 
around a mean of zero, but also a shift in the mean to a positive value.  
Light-paths are attracted by regions of overdensity and repelled by regions of 
underdensity.  This means that the column density of mass seen by the observer 
is larger on average than what would be expected using unperturbed 
light-paths.  The predominantly positive second derivatives of the potential in overdense regions 
produces a shear between light paths which acts to magnify images.
At the same time, the average shear between light-paths is, to a lesser extent, reduced by the 
increase in the density of light paths in overdense regions.  The net result is that objects on  
surfaces of equal redshift or cosmological time will on average appear larger than in an unperturbed 
universe.  The apparent violation of flux conservation can be resolved by realizing that the area of a 
surface of constant redshift is smaller when light-paths are perturbed.  In angular size coordinates, 
light travels ``slower'' in regions of low potential.

The other and more important aim of this paper is to show that after lensing the CMB power spectrum 
will be enhanced over the unlensed power spectrum at small angular scales or large $\ell$.  Power is 
transferred upward in $\ell$ in the damping tail.  This result 
is independent of the existence of a nonzero mean magnification.  The paper is organized as 
follows: In the next section we introduce the formalism used to calculate the lensing 
effects of LSS.  In section 3 it is shown how lensing will change a generic CMB power 
spectrum.  In section 4 the formalism is applied to some specific cosmological models.

\section{Calculating the Magnification}

Throughout this paper, the Universe is assumed to have Robertson-Walker 
geometry together with small  fluctuations.  This implies that the 
density fluctuations are isotropic and the universe is  homogeneous on the average.  We also assume 
that the lensing is weak so that there are not multiple images of a single source.  It can be 
shown without difficulty that the cross-section of regions with densities over the critical 
density required to produce multiple images is rather small, so that they should not play an 
important role in the statistical properties of lensing over large regions of sky \cite{sch92},
\cite{koch95}.

In the longitudinal gauge with conformal time, the metric takes the form
\begin{equation}
ds^2 = a(\tau)^2 (-(1+2\phi )d\tau^2 +(1-2\phi ) dx^2); \ \ 
dx^2 = dr^2 + g(r)^2 (d^2 \theta + \sin^2(\theta)d^2 \phi),
\end{equation}
where $\phi \ll 1$ and $g(r) = \{ R\sinh(r/R),r,R\sin(r/R) \}$ for the open, flat and closed 
global geometries respectively.  The curvature scale is $R=|H_o \sqrt{1-\Omega-\Omega_{\Lambda}}|^{-1}$.  Because Maxwell's equations are conformally 
invariant, for the purpose of finding light-paths the expansion of the universe can be 
ignored as long as conformal time is used.  In general, light follows a geodesic that 
is a solution to:
\begin{equation}
\frac{d}{d\lambda} g_{\mu \nu} \frac{d x^{\nu}(\lambda)}{d\lambda} = \frac{1}{2} g_{\alpha \beta,\mu} \frac{d x^{\alpha}(\lambda)}{d\lambda} \frac{dx^{\beta}(\lambda)}{d\lambda}
; \qquad p^0 = \frac{d\tau}{d\lambda} = a(t)^{-1}\frac{dt}{d\lambda }.
\end{equation}
Choosing $\lambda = \tau$ by normalizing $p^0$ and taking the unperturbed 
path to be the $r$-axis, the evolution equation to first order in the potential 
$\phi$ becomes 
\begin{equation}
\frac{d^2 \delta \theta}{d\tau^2} = -2 g(r)^{-2}\phi_{,\theta}=-2 g(r)^{-1} \phi_{,\perp}.
\end{equation}
Since $\tau = -r + \tau_o$ to first order in $\phi$ this equation can be 
solved as a function of $r$:
\begin{equation}
\delta \theta_{i}(r) = \frac{\delta x_{i}(r)}{g(r)} = -\frac{2}{g(r)}\int^r_0 dr' g(r-r')\phi_{,i}(r').
\label{deltx}
\end{equation}
This must be evaluated along the path that the light bundle has followed.  
The first order effects arise from evaluating it along the unperturbed path.  
To find the correction due to the perturbation of the path,   we expand the 
potential to first order:

\begin{equation}
\delta \theta_{i}(r) = \frac{\delta x_{i}(r)}{g(r)} = -\frac{2}{g(r)}\int^r_0 dr' g(r-r') \left( \phi_{,i}(r') + \delta x^k(r') \phi_{,ik}(r') \right)
\label{expand}
\end{equation}
where the potential is now evaluated along the unperturbed path.  Repeated indices are 
summed over the two components perpendicular to this path.  Likewise the 
$\delta x(r')$ inside the integral can be approximated by the first order deflection 
calculated from equation~(\ref{deltx}) evaluated along the unperturbed path.
The shear tensor which measures the distortion and expansion of an infinitesimally 
thin beam is then
\begin{eqnarray}
\Phi_{ij} & \equiv & \frac{\partial \delta \theta_i}{\partial \theta_j} =  \Phi^o_{ij} + \Delta \Phi_{ij} = \frac{-2}{g(r)}\int^r_0 dr' g(r')g(r-r')\phi_{,ij}(r') 
\label{deltphi} \\
& & + \frac{4}{g(r)}\int^r_0 dr' \int^{r'}_0 dr'' g(r-r') g(r'-r'') 
\left[ g(r')\phi_{,k}(r'') \phi_{,ijk}(r') + g(r'')\phi_{,jk}(r'') \phi_{,ik}(r') \right]. \nonumber
\end{eqnarray}
In general this expansion is not justified for fluctuations of all scales.  
However it can be shown by explicit calculation that higher order terms are quite small in 
realistic models.  If we assume that the relevant scales are much smaller than the curvature scale 
we can Fourier decompose the potential,
\begin{equation}
\phi (r) = \int \frac{d^3 k}{(2\pi )^3} {\tilde \phi}(k,\tau=\tau_o-r) e^{-ikx}e^{-i(rk_r 
+ g(r)\overline{\theta}\cdot k_{\perp})}.
\label{fourier}
\end{equation}
In this section we assume that the angles involved are small enough that a local Cartesian 
coordinate system, $\overline\theta$, can be set up with the usual inner product.

The average value of the shear tensor can be found by substituting 
equation~(\ref{fourier}) into equation~(\ref{deltphi}) and using the assumption 
that the Fourier components are uncorrelated, i.e. $\langle{\tilde \phi}(k,\tau){\tilde 
\phi}(k',\tau)^*\rangle=(2\pi)^3 P_{\phi}(k,\tau) \delta(k-k')$.  In the case of the linear 
evolution of the potential fluctuations in a universe dominated by nonrelativistic 
matter, the time dependence of the potentials can be factored out of its Fourier components, 
${\tilde\phi}(k,\tau)=D(\tau){\tilde\phi}(k)$.  In this case,
\begin{eqnarray}
\left\langle \Phi_{ij} \right\rangle & = & -2 \delta_{ij} \int \frac{d^3 k}{(2\pi )^3} 
k_{\perp}^2 k_{\perp}^2 P_{\phi}(k) W(r,r k_r,g(r) \overline{\theta}\cdot k_{\perp}),
\label{phipot} \\
 W(r,r k_r,g(r) \overline{\theta}\cdot k_{\perp}) & = & \frac{1}{g(r)} 
\int^r_o dr'\int^{r'}_0 dr'' D(\tau')D(\tau'') \label{wind} \\
& & \times g(r-r')g(r'-r'')[g(r')-g(r'')]e^{-i[(r'-r'')k_r 
+\left(g(r')-g(r'')\right){\overline\theta}\cdot k_{\perp}]}. \nonumber
\end{eqnarray}
Equation (\ref{wind}) can be interpreted as consisting of two contributions.  The term with $g(r')$ 
is due to the average potential, as sampled by the light paths, being below average.  The $g(r'')$ term 
results from the density of light paths being higher in areas of low potential.  The reduced 
separation between light paths makes them converge less rapidly.  The second term almost cancels the 
first term because in popular models the $k$ values that contribute most are large enough that the 
oscillations of the exponential restrict $r'-r''$ to be small.  It appears that the coherence length 
of structure is small enough to make this magnification negligible.

 The time enters into these 
calculations because it is a function of the radial coordinate that parameterizes the 
light path.  All the significant quantities calculated in this section, such as  the second term in 
equation (\ref{wind}), contain two integrations over this 
parameter.  However, when $|r'-r''|(= |\tau'-\tau''|)$ is large, larger then some ill-defined 
``coherence length'', the potential fluctuations 
at these two points are uncorrelated and do not contribute significantly to the 
integrals.  If the potential changes slowly enough it will not change significantly 
in the time it takes light to travel one ``coherence length'' and we can 
take  $\langle{\tilde \phi}(k,\tau){\tilde \phi}(k',\tau')^*\rangle=(2\pi)^3 
D\left({\overline\tau}=(\tau+\tau')/2\right)^2 P_{\phi}(k) \delta(k-k')$.  We will call 
this the average time assumption.  It can be avoided at the expense of complicating the 
evaluation of the integrals.\footnote{If a
significant amount of hot dark matter exists or there has been substantial nonlinear 
evolution, the factorization of ${\tilde\phi}$ into $k$- and $\tau$-dependent parts 
will not be possible.  In these models the whole power spectrum must be kept within the $r$ integrals 
which in general must be evaluated numerically.}  
These complications of time evolution are largely avoided in the flat-CDM model 
because the potential is time-independent in linear theory.

The quantity of interest for applications to the CMB is the difference in the 
deflections of light paths that are observed to have an angular separation of 
$\overline{s}$ on our sky.  This can be found by integrating the shear tensor
\begin{equation}
\beta_i(s) \equiv \int_{-s/2}^{s/2} \Phi_{ij}(\theta) d\theta_j.
\label{betadeff}
\end{equation}
$\beta(s)$ has both a small average due to second order terms and a variance which is dominated by 
first order terms.  Combining equations 
(\ref{betadeff}), (\ref{fourier}) and (\ref{deltphi}),
\begin{eqnarray}
\left\langle \beta_{\parallel,\perp}(s)^2 \right\rangle & 
= \frac{\pi}{2} \left( \frac{4}{g(r)} \right)^2 
\int_0^r dr' \int_0^r dr''\int \frac{dk_r dk_{\perp}}{(2\pi )^3} k^3_{\perp} 
P_{\phi}(k,{\overline\tau}) g(r-r')g(r-r'')e^{-ik_r (r'-r'')} 
\label{betabeta} \\
 & \times \left\{ J_o[w_{-}] - J_o[w_{+}] \pm \left( J_1[w_{-}]/w_{-} - J_1[w_{+}]/w_{+} \right)
\right\},
\nonumber
\end{eqnarray}
where $w_{-}=k_{\perp}s g(r'-r'')/2$ and $w_{+}= k_{\perp}s g(r'+r'')/2$.  The plus sign in the 
second line is for $\beta_{\parallel}$, the component parallel to $\overline{s}$, and the minus is for 
the perpendicular component, $\beta_{\perp}$.  In 
these coordinates the cross terms vanish.  To reduce the numerical work necessary to integrate these 
oscillatory integrands it is useful to make an approximation.  This approximation can be understood 
by first changing variables from $\{r',r''\}$ to $\{\overline{r}=(r'+r'')/2,y=r'-r''\}$.  When $y$ 
is large the fluctuations in $\phi$ will add incoherently.  We assume that the 
coherence length is small enough that $g(r-r') \simeq g(r-r'') \simeq 
g(r-\overline{r})$ and that $J_o(k_{\perp}s y/2) \simeq 1$ for all relevant $k_{\perp}$ and $s$.  Then the $y$ integration of the exponential can be done giving 
$2 r j_o(rk_r)$.  This spherical Bessel function suppresses the contribution of modes 
with $k_r$ much greater than $1/r$.  It is usually the case in lensing situations that the 
peak in $P_{\phi}(k)$ is at a $k\gg 1/r$ so that everything but $j_o(rk)$ can be brought 
out of the $k_r$ integral and $k_{\perp}$ replaced with $k$.  In other words, the 
modes that contribute most are nearly perpendicular to the line of sight.  The result 
is
\begin{equation}
\left\langle \beta_{\parallel,\perp}(s)^2 \right\rangle  \simeq 
\left( \frac{2}{g(r)} \right)^2 \int_0^r d\overline{r} \int \frac{dk}{2 \pi} k^3 
P_{\phi}(k,{\overline\tau})  g(r-\overline{r})^2 \left\{ 1- J_o[k s g(\overline{r})] 
\pm J_1[k s g(\overline{r})]/k s g(\overline{r}) \right\}. 
\label{beta2}
\end{equation}

This is equivalent to result that is found in \cite{seljak} using the Fourier space Limber's 
equation derived by Kaiser \cite{kais92}.  We find explicitly that for popular models, equation 
(\ref{beta2}) estimates equation (\ref{betabeta}) quite well for both large and small s, and 
appears to do well for intermediate values.

\section{The CMB Power Spectrum}

We wish to calculate the effect gravitational lensing should have on fluctuations in the 
CMB.  Lensing will not change the temperature or surface 
brightness of the CMB, but it will change the size and shape of features. In this paper we 
restrict ourselves to the CMB power spectrum.  In \cite{seljak} a relation between 
the lensed and unlensed power spectra was derived by replacing the spherical harmonic transformation 
of the CMB fluctuations with its Fourier transform.  This is a good approximation if $\ell \gg 1$ 
and $\ell s \ll 1$ for all the relevant scales of $s$ and $\ell$.  In the appendix we derive a relation 
that avoids the second assumption by remaining in spherical harmonic space.  We also retain the 
anisotropic contributions that are dropped in \cite{seljak}.  We find that for the models tested the 
two methods agree very well if the anisotropic contributions are included in both cases.  Given that the 
two methods of calculation give the same results we prefer ours because we find it to be the faster one 
in our implementations.

The transformation of the power spectrum derived in the appendix is:
\begin{eqnarray}
C_\ell^{ob} & \simeq & \sum_{\ell'=0}^{\infty} C_{\ell'} \frac{2\ell'+1}{2} \int_0^{\pi} ds \sin(s) 
P_\ell[\cos(s)]\left\{ e^{-\ell'^2\langle \beta_{\parallel}(s)^2 \rangle/2} P_{\ell'}[x] \right.
\label{Ctrans} \\
& & \left. +\frac{1}{2}\left[ \langle \beta_{\parallel}(s)^2 \rangle - \langle \beta_\perp(s)^2 \rangle \right] P'_{\ell'}[x] 
\right\}_{x=\cos(s)} \nonumber
\end{eqnarray}
where $C_\ell^{ob}$ is the observed power spectrum.

\section{Applications}

For the purpose of making some quantitative predictions, we adopt a flat, $\Lambda = 0$, 
cold dark matter cosmological model with adiabatic initial density perturbations.  We use the 
CDM linear power spectrum given by \cite{bard86} with the correction for finite baryon density
given by \cite{sugy95}:
\begin{eqnarray}
&P(k)=Ak^n T(k e^{2\Omega_b}/h^2)^2 &
\\
&T(q)=\frac{\ln(1+2.34q)}{2.34q}\left[ 1+3.89q+(16.1q)^2+(5.46q)^3+(6.71q)^4 \right]^{-1/4}&
\end{eqnarray}
This is related to the potential power spectrum by Poisson's equation.  The Hubble parameter, 
$H_o=100 h \mbox{ km s}^{-1}\mbox{Mpc}^{-1}$ and the COBE normalization is $A=6.6 \times 10^{5} h^{-(3+n)} \mbox{ Mpc}^4$ \cite{benn96}.  The nonlinear evolution of the power spectrum is calculated using the fitting formulae of \cite{peac96}.

\begin{figure}
\epsscale{0.55}
\plotone{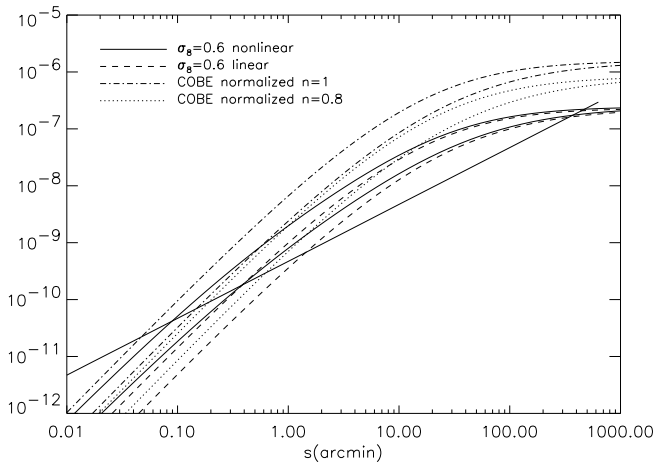}
\caption[stuff]{\small The first two moments of $\beta(s)$ for some CDM models are given here.  The straight 
line is $\langle \beta_\parallel(s) \rangle$ in a COBE normalized model.  The two curves of each type 
are $\langle \beta_\parallel(s) \rangle$ (the larger) and $\langle \beta_\perp(s) \rangle$ (the 
smaller).  The COBE normalized models have only linear evolution of the matter power spectrum.  All 
the models have $h=0.6$ and $\sigma_8$ is the rms density fluctuation in a sphere of radius 
$8 h^{-1}$ Mpc.}
\label{betafig}
\end{figure}

The second moments of $\beta_{\parallel}(s)$ and $\beta_{\perp}(s)$ are given in figure~(\ref{betafig}) 
along with an example of the 
first moment.  It can be seen that at small angles 
$\langle \beta(s)^2 \rangle$ is approximately proportional to $s^2$.  At large separations the 
deflections of the two 
paths are uncorrelated, and  so $\beta_{\parallel}(s) \simeq \beta_{\perp}(s) \simeq \langle 
\delta(\theta)^2 \rangle$ which is independent of $s$.  This constant could also be calculated using 
equation~(\ref{deltx}).  In general increasing the Hubble parameter increases lensing effects because 
it shifts power down to smaller physical scales.

\begin{figure}
\epsscale{0.75}
\plotone{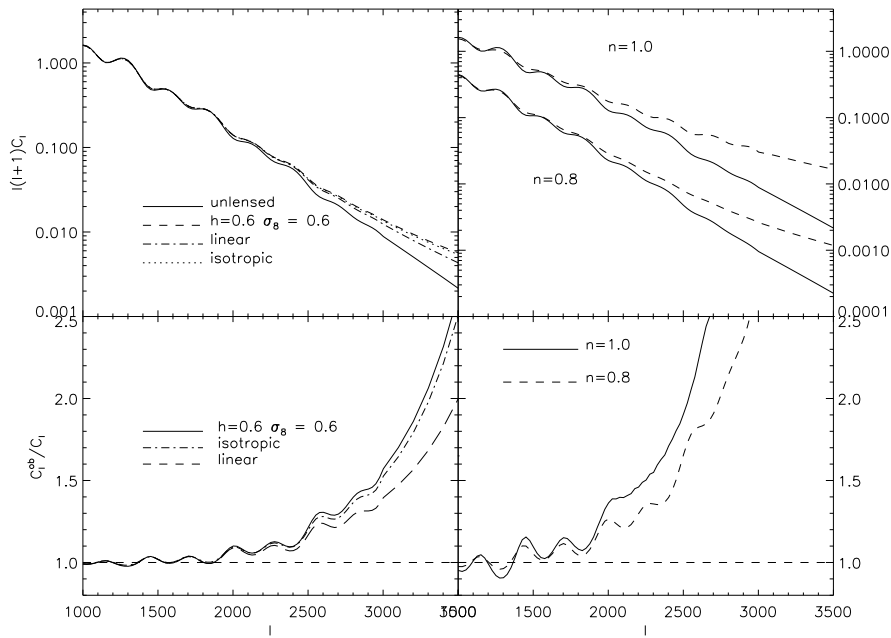}
\caption[stuff]{\small The lensed CMB power spectrum for a CDM model with $h=0.6$, $\Omega_o=1$ and 
$\Omega_b=0.04$:  The top panels are the lensed and unlensed spectra and on the bottom are their 
ratios.  The left panels show $n=1.0$, $\sigma_8=0.6$ normalized models.  The 
linear matter power spectrum evolution and the isotropic lensing approximations are shown.
The right panels show the effect of tilting the power spectrum with COBE normalization and linear 
evolution.  In the top right panel the solid curves are the unlensed spectra.}
\label{spec1}
\end{figure}

Figure~(\ref{spec1}) shows the effect lensing has on the CMB power spectrum calculated 
using equation~(\ref{Ctrans}).  In doing this calculation for large $\ell$ one must take 
care that the Legendre polynomials and the integration are being accurately 
calculated.  The unlensed spectra are produced using Seljak \& 
Zaldarriaga's code, \cite{seljak96}, up to $\ell = 3000$.  Beyond this point a fitting formula for 
the damping tail $\ell(\ell+1)C_\ell \propto \ell^{(n-1)} \exp[-2(\ell /\ell_D)^m]$ is used where $m$ 
and $\ell_D$ are given in \cite{hu96}.  The plots of $C_\ell^{ob} / C_\ell$ shows not only that the 
fractional changes in the $C_{\ell}$'s are substantial at large $\ell$, but also that the damping tail 
of the observed and the intrinsic, unlensed spectra will have different functional forms.
The calculation is also done with the $J_1$ terms dropped from (\ref{beta2}) so that the 
lensing is isotropic, $\langle \beta_{\parallel}(s)^2 \rangle=\langle \beta_{\perp}(s)^2 \rangle $.
The results reproduces the ones we get from equation~(A8) in 
\cite{seljak}.  When $n$, the spectral index of the primordial power spectrum, is reduced the 
steepening of the damping tail partially compensates for the weaker lensing and results in a 
comparatively small change in $C_\ell^{ob}/C_ell$.  Including nonlinear structure formation is found 
to increase the lensing effect significantly for $\ell \gtrsim 2000$.  The range of $\ell$-space that 
contributes in (\ref{Ctrans}) is surprisingly large.  At $\ell=3000$ about $30\%$ of the change in power 
is due to contributions from $\ell<1000$ and $8\%$ is from $\ell<500$, in the $\sigma_8=0.6$ model.  
In this sense lensing can be thought of as transferring power from the acoustic peaks to the damping 
tail in such a way that the variance is conserved.

The parameters $m$ and $\ell_D$ are dependent on $h^2\Omega_o$, $h^2\Omega_b$, $h^2\Omega_{\Lambda}$, 
and the average temperature.  Hu \& White~\cite{hu96} have proposed that measuring $m$ and $\ell_D$ 
would be a good 
way of determining these density parameters.  However lensing will cause the spectrum to 
fall less rapidly than expected.  
If we fit a damping curve to the $1000 < \ell < 3000$ part of the $\sigma_8=0.6$-nonlinear lensed 
spectrum in figure~(\ref{spec1}) $ m $ and $\ell_D$ are changed by $19\%$ and $-17\%$ from the unlensed 
spectrum.  If measured in this way lensing will cause 
the angular size of the damping scale at the time of decoupling, $\ell_D^{-1}$, to be overestimated.  
Likewise a small value for $m$ would mean that the thickness of the surface of last scattering would be 
overestimated.  The change in $m$ is especially significant since it is otherwise quite a weak 
function of cosmological parameters.  Within the acceptable range of parameter space $m$ changes by 
only about $10\%$ in the unlensed spectrum.

\section{Conclusion}

We have shown that gravitational lensing can have a significant effect on the CMB power 
spectrum at small angular scales.  The mean magnification will probably be too small to 
detect, but variations in the magnification will cause the damping tail to decrease less 
rapidly with increasing $\ell$.  We have also found that nonlinear structure formation and 
anisotropic contributions the transformation of the power spectrum are important at large $\ell$.
Acoustic peaks at large $\ell$ may be smoothed to such an extent that 
they are unidentifiable.  The effects of lensing can be removed from the spectrum, 
but a model for both the lensing potential and the unlensed CMB power spectrum must be 
assumed.  In addition the transformation of the power spectrum is nonlinear although it 
seems well behaved.  This increases 
the amount of potential information in the damping tail, but makes the interpretation of 
future small-scale observations more ambiguous.  Perhaps the power spectrum of density 
perturbations will be more tightly constrained by other means.

In this paper we have only shown results for flat cosmological models with no cosmological constant, 
$\Lambda$.  Lensing effects will be somewhat smaller in both low density and $\Lambda$ models, 
because of the appearance of the mass density in Poisson's equation, $P_{\phi}(k,\tau)=9 
a(\tau)^{-2}\Omega^2_o \mbox{H}_o^4 k^{-4} P(k,\tau)/4$.  This factor overcompensates for 
the increase in path length and growth in fluctuations with lookback time \cite{seljak}.  With 
reasonable values for $\Omega_o$ and $\Lambda$ the change in the spectrum is still 
significant.

Interferometers are under construction that will be capable of probing the predicted CMB fluctuations 
from 150 to 3500 in $\ell$.  The window size in $\ell$-space for an 
interferometer is $\sim 2\pi D$ where $D$ is the diameter of the dishes in units of the wavelength.  
The proposed instruments will operate at around $30$GHz and have dish diameters of tens of 
centimeters so an $\ell$-space resolution of $100$ should be achievable.  This should allow for many 
independent measurements of the rate at which the tail falls with $\ell$.  In addition, mosaicing over 
the sky can further narrow the window.  These experiments will use multiple frequencies so that 
foregrounds such as the Sunyaev-Zel'dovich effect can be removed.  We have shown that lensing 
corrections increase the amplitude by a factor of $\sim 2$ or more above $\ell\sim 3000$ in flat CDM 
models with Hubble constants in the observed range.  With $\ell$-space 
windows in the above range any experiment that is capable of detecting the unlensed spectrum at these 
high $\ell$'s will be measurably affected by lensing.  It will then be essential to include the 
lensing contribution to the CMB fluctuations in order to utilize the tail beyond $\ell\sim 2000$ for 
cosmological parameter estimation.

\acknowledgments{The authors would like to thank U. Seljak, E. Martinez-Gonzalez, J.-L. Sanz and R. 
Bar-Kana for helpful comments on an early version of this paper.  This research has been supported in 
part by a grant from NASA.}

\appendix

\section{Appendix}

We start by expressing the observed temperature fluctuations, $I_o(\overline\theta)\equiv 
(T(\overline\theta)-\langle T \rangle)/\langle T \rangle$, in terms of the spherical 
harmonic expansion of the intrinsic, unlensed fluctuations,
\begin{equation}
I_o(\theta,\phi)=\sum_{\ell =0}^\infty \sum_{m=-\ell}^\ell a_{\ell m} Y_{\ell}^{m}(\theta',\phi')
\end{equation}
where $\{\theta',\phi'\}$ is the position from which the light that is observed at  position 
$\{\theta,\phi\}$ would have come from if there were no lensing.  The two-point 
correlation function at an angular lag of $s$ can be found with the help of 
$\langle a_{\ell m} a_{\ell' m'}^* \rangle = \delta_{\ell\ell'} \delta_{m m'} C_{\ell}$ 
and the addition theorem for spherical harmonics,
\begin{equation}
\xi(s) \equiv \langle I_o(\theta_1,\phi_1) I_o(\theta_2,\phi_2) \rangle = \sum_{\ell=0}^\infty \frac{(2\ell +1)}{4\pi}
C_\ell \langle P_\ell[\cos(|\overline{s}+\overline\beta(s)|)] \rangle,
\end{equation}
where the $|\overline{s}+\overline\beta(s)|$ is the angle that would separate the two points on 
the surface of last scattering if there where no lensing.  It has been assumed that the intrinsic 
fluctuations in the CMB are uncorrelated with the fluctuations in the potential that contribute to 
the lensing.  This is a very good approximation in realistic models.  Because $\beta(0)=0$, 
the variance of the CMB is unchanged by lensing, i.e. the quantity $\sum_{\ell=0}^{\infty} 
(2\ell+1)C_{\ell}$ is conserved.

Now the correlation function 
can be transformed back into the observed power spectrum using $C_\ell^{ob} = (2\pi) \int_0^\pi ds 
\sin(s) \xi(s) P_\ell[\cos(s)]$.  To express the $C^{ob}_{\ell}$ in terms of moments of $\beta(s)$ 
instead of the average of Legendre functions we expand these functions in a power series
\begin{eqnarray}
 \langle P_\ell[\cos(|\overline{s}+\overline\beta(s)|)] \rangle & \simeq &\sum^{\infty}_{n=0}  \frac{(-1)^n\sin(s)^n}{n!}  \langle \beta_\parallel(s)^n \rangle P^{(n)}_{\ell}[x] \label{pser} \\
& & +\frac{1}{2} \left[ \cos(s)\langle \beta_\parallel(s)^2 \rangle - \frac{\sin(s)}{s} \langle \beta_\bot(s)^2 \rangle  \right] P'_{\ell'}[x]. \nonumber
\end{eqnarray}
where the prime and $(n)$ superscript represent the first and nth derivative with respect to $x=\cos(s)$. 
 In each $n$-term, except $n=1$, only the terms with the lowest powers of $\beta(s)$ are kept.  Since the $n$th 
derivative is of order $\ell$ larger than the $(n-1)$th derivative and $\ell$ is large, the higher 
order terms are not important.  Because the component of $\beta(s)$ parallel to $\overline{s}$, 
$\beta_\parallel(s)$ and the component perpendicular to $\overline{s}$, $\beta_\bot(s)$, are uncorrelated 
their cross terms are of order 
$\langle \beta(s)^2 \rangle^{2n}$ for small $s$ and do not contribute significantly.  For $n=1$, 
higher order terms are kept because in practice they contribute significantly to the transformation 
of the power spectrum.

Only the first term in the $\beta_\perp(s)$ has been kept.  A noticeable improvement in accuracy 
can be made by including higher order terms in the $\beta_\parallel(s)$ series.  This is because at 
some point $\ell$ becomes of order $\beta(s)^{-1}$ and $P_{\ell}[x]$ oscillates with a period 
of order $\beta(s)$.  This problem can be effectively circumvented if $\left( \beta_\parallel(s)-\langle 
\beta_\parallel(s) \rangle \right)$ is a Gaussian random variable.  Since $\beta_\parallel(s)$ is 
the result of a radial integral over many coherence lengths the central limit theorem supports 
the assumption of Gaussianity even if the 
potential field is weakly non-Gaussian.  In practice $\langle \beta(s) \rangle$ is small enough 
that we need only keep its first term in equation~(\ref{pser}).  For the other terms Gaussianity requires $\langle \beta(s)^{2m} \rangle \simeq 
(2m)! \langle 
\beta(s)^{2} \rangle^m/2^m m!$ and for large $\ell$, $P_{\ell}^{(2m)}[x] \simeq (-1)^m \ell^{2m} 
P_{\ell}[x]/(1-x^2)^{m}$ which can be easily shown directly from Legendre's equation.  
As will be seen the lensing has significant effects only at large $\ell$.
With these simplifications 
\begin{eqnarray}
C_\ell^{ob} & \simeq & \sum_{\ell'=0}^{\infty} C_{\ell'} \frac{2\ell'+1}{2} \int_0^{\pi} ds \sin(s) 
P_\ell[\cos(s)]\left\{ e^{-\ell'^2\langle \beta_\parallel(s)^2 \rangle/2} P_{\ell'}[x] \right.
\label{CtranA}\\
& & \left. +\frac{1}{2}\left[  \cos(s)\langle \beta_\parallel(s)^2 \rangle - \frac{\sin(s)}{s} \langle \beta_\bot(s)^2 \rangle \right]  P'_{\ell'}[x]-\langle \beta_\parallel(s) \rangle \sin(s) P'_{\ell'}[x] \right\}_{x=\cos(s)} \nonumber
\end{eqnarray}
It is clear from this expression that the change in the power spectrum will become significant when 
$\ell^2 \langle \beta(s)^2 \rangle$ is significant.  Neglecting the first moment of $\beta(s)$ and 
taking the small angle limit of the $P'_{\ell'}[x]$ term, which is a good approximation 
in practice, gives equation~(\ref{Ctrans}).

\end{document}